\documentclass[twocolumn,showpacs,preprintnumbers,amsmath,amssymb]{revtex4}

\usepackage{graphicx}
\usepackage{dcolumn}
\usepackage{bm}

\begin{document}

\title{Strong-coupling dynamics of a multi-cellular chemotactic system}
\author{R. Grima $^{1,2}$}

\affiliation{$^{1}$ Department of Physics and Astronomy,
Arizona State University, Tempe, AZ 85284 \\
$^{2}$ Biocomplexity Institute, Department of Physics, Indiana
University, Bloomington, IN 47405}

\begin{abstract}
Chemical signaling is one of the ubiquitous mechanisms by which
inter-cellular communication takes place at the microscopic level,
particularly via chemotaxis. Such multi-cellular systems are
popularly studied using continuum, mean-field equations. In this
letter we study a stochastic model of chemotactic signaling. The
Langevin formalism of the model makes it amenable to calculation via
non-perturbative analysis, which enables a quantification of the
effect of fluctuations on both the weak and strongly-coupled
biological dynamics. In particular we show that the (i)
self-localization due to auto-chemotaxis is impossible. (ii) when
aggregation occurs, the aggregate performs a random walk with a
renormalized diffusion coefficient $D_R \propto \epsilon^{-2}
N^{-3}$. (iii) the stochastic model exhibits sharp transitions in
cell motile behavior for negative chemotaxis, behavior which has no
parallel in the mean-field Keller-Segel equations.

\end{abstract} \vspace{2mm} 

\maketitle

The study of biological systems through modeling is a promising
endeavor to understand or throw light on the macroscopic complexity
originating from the microscopic cellular interactions common to all
living organisms. At the microscopic level, cells interact with each
other through various means, principally via local short-range
forces such as adhesion and through long-range forces mediated via
chemical signals. In many cases, cells do not just respond to
chemical signals but are actively involved in their production also.
This signal feedback leads to intricate inter-cellular
communication, which is the main mechanism behind the emergence of
the observed complex behavior of multi-cellular systems. An
important aspect of the feedback mechanism is that the cells'
dynamics are typically dominated by long range spatio-temporal
correlations. Modeling has traditionally been approached through the
construction of coupled partial differential equations, describing
the evolution of a density field $\rho$, representing the number
density of cells. Many of these models are variants of the
Keller-Segel equations \cite{Keller}. Recently it has been shown
that the derivation of the latter equations from a microscopic,
stochastic Langevin model of interacting cells, is achieved by
neglecting cell-cell correlations \cite{Newman}; indeed this
verifies the hypothesis that Keller-Segel variants are mean-field
type models i.e. they are applicable to modeling biological
situations in which the cell number density is sufficiently large.
This statement is however qualitative; it is not clear what are the
similarities and differences predicted by the stochastic models and
their deterministic counterparts.

In this letter we study a stochastic model of chemotactic signaling;
this being an individual-based model of cells interacting via
long-range chemical signals and actively responding to such signals
via chemotaxis. Such models have been previously studied by a number
of authors (see for example \cite{Othmer}, \cite{Stevens},
\cite{Jiang}, \cite{Hadeler}, \cite{Merks} ). We shall show that it
is possible to gain an understanding of the cells'
strongly-correlated dynamics by means of a non-perturbative analysis
applied directly on the Langevin equation formalism of the model.
This will give us an analytical quantitative way of comparing the
stochastic and deterministic models. It is to be emphasized that the
non-perturbative nature of the analysis method will enable us to
obtain insight, otherwise not obtainable via the conventional
perturbative approach \cite{Newman} or through analysis of the
corresponding mean-field type equations. The system we shall analyze
consists of $N$ chemotactic cells which are constantly secreting a
chemical (whose concentration is denoted by $\phi$) and which
respond to the local chemical gradient by either moving up the
gradient (positive chemotaxis) or down the gradient (negative
chemotaxis). The latter leads to dispersion whereas the former
effect leads to aggregation. Such mechanisms are common to many
organisms including amoeba, myxobacteria, leucocytes, and germ
cells. We shall first treat the case of a single self-interacting
cell, then extend it to the multi-cellular case. The equations
defining the single-cell stochastic model are \cite{Newman}:
\begin{align}
\label{e1}
{\dot {\bf x}_{c}}(t) &= {\bf \xi}(t) + \kappa \alpha \nabla \phi({\bf x}_{c},t) , \\
\label{e2}
\partial _{t} \phi({\bf x},t) &= D_{1}\nabla ^{2} \phi({\bf x},t) - \lambda \phi({\bf x},t)
+\beta \delta ({\bf x}-{\bf x}_{c}(t)).
\end{align}
Eq.{\ref{e1}} is a Langevin equation describing the motion of a cell
whose position at time $t$ is denoted as ${\bf x}_{c}(t)$. The
stochastic variable $\xi$ is white noise defined through $\langle
\xi^{a}(t) \rangle = 0$ and $\langle \xi^{a}(t)\xi^{b}(t') \rangle =
2 D_{0}\delta_{a,b}\delta(t-t')$ where $a$ and $b$ refer to the
spatial components of the noise vectors. In the absence of a
chemical gradient, the cell performs a pure random walk
characterized by a diffusion coefficient $D_0$. In the presence of a
chemical gradient, the cell has a velocity $\kappa \alpha \nabla
\phi$ superimposed on the random walk, where $\alpha$ is a positive
constant typifying the strength of chemotaxis and $\kappa$ is a
constant which can take the values $-1$ (negative chemotaxis) or $1$
(positive chemotaxis). The overall effect is a random walk biased in
the direction of increasing chemical concentration ($\kappa = 1$) or
in the direction of decreasing chemical concentration ($\kappa =
-1$). Eq.{\ref{e2}} is a reaction-diffusion equation describing the
chemical dynamics. The chemical diffuses with diffusion coefficient
$D_1$, decays in solution at a rate $\lambda$ and is secreted by the
cell at a rate $\beta$. The feedback mechanism is what makes this
problem non-trivial. The cell constantly modifies its environment
through its continuous chemical secretion and simultaneously reacts
to its environment via chemotactic sensing and directed motion. For
positive chemotaxis, the net effect of the two coupled equations
gives rise to a random walk having a larger probability of visiting
spatial areas which it has previously visited than of visiting
previously unexplored regions. For negative chemotaxis, the opposite
situation occurs: the walker is ``repelled'' away from regions which
it has previously visited. The self-interaction of a cell will be
referred to as auto-chemotaxis.

The strong non-Markovian nature of the dynamics is what makes this
and similar problems (involving self-interacting random walks)
difficult to analyze. In this letter we introduce a non-perturbative
method to explore the strong-coupling aspects of the theory. Unlike
perturbation theory in the coupling parameter $\epsilon = \alpha
\beta$ \cite{Newman}, this method can be applied directly to the
Langevin formulation of the model i.e. the analysis bypasses the
conventional derivation of the equations of motion for the single
and multi-cell probability distributions. Integrating the chemical
equation Eq.{\ref{e2}}, assuming that there is no chemical initially
$\phi(\mathbf {x},0)=0$, one finds an expression for the local
chemical gradient sensed by the cell at time $t$:
\begin{multline}
\label{e5} \nabla \phi= -\frac{\beta}{2} (4\pi t)^{-d/2}
D_{1}^{-(1+d/2)} \int\limits_{\tau / t}^{1} du \frac{[\mathbf
{x}_c(t)-\mathbf{x}_c(t-ut)]}{u^{1+d/2}} \\ \times \exp \Biggl[
-\lambda t u - \frac{ [x_c(t)-x_c(t-ut)]^{2}}{4 D_1 t u} \Biggr],
\end{multline}
where $d$ is the dimensionality of space and $\tau$ is a refractory
period i.e a period of time in which the cell is not sensitive to
chemical signals, introducing an effective time delay between signal
emission and signal transduction. Another way of stating this is
that the cell at time $t$ senses the local gradient due to chemical
production in the period $t' \in (0,t-\tau)$. Such an effect is a
common feature of many chemotactic cells \cite{Bray}. The
introduction of $\tau$ also regularizes the integral in
Eq.{\ref{e5}}. Although it is in general impossible to solve this
integral, since this requires full knowledge of all previous cell
positions, in the asymptotic limit $t \gg 1/\lambda$ the integral is
dominated by small $u$ \cite{Murray}. It may therefore be simplified
by use of the approximation $\mathbf {x}_c(t)-\mathbf{x}_c(t-ut)
 \simeq ut {\dot {\bf x}_{c}} (t)$. We further introduce two
 convenient variables:
$\gamma = 2 \epsilon \pi^{-d/2} (4D_1)^{-(1+d/2)}$ and $\lambda' =
\lambda + \frac{[{\dot {x}_{c}}(t)]^{2}}{4 D_1}$. Substituting the
resulting expression for the chemical gradient in the Langevin
equation for the cell we get
\begin{equation}
\label{e6} {\dot {\bf x}_{c}}(t) = {\bf \xi}(t) - \kappa {\dot {\bf
x}_{c}}(t) \Biggl[ t^{1-d/2} \gamma \int\limits_{\tau / t}^{1} du
\frac {e ^{ {-\lambda' t u}}} {u^{d/2}} \Biggr].
\end{equation}
Thus we have showed that the long time dynamics of a
self-interacting chemotactic cell can be described by a modified
Langevin type equation. The explicit computation of the integral on
the R.H.S of Eq.(\ref{e6}) leads to the following expressions for
$d=1,2$ and 3 respectively
\begin{align}
\label{e7} {\dot {\bf x}_{c}}& = {\bf \xi} - \kappa \frac{\epsilon
(1-erf \sqrt{\lambda \tau}) \dot {\bf x}_{c}}{4 D_1^{3/2}
\sqrt{\lambda} }
\biggl({1 + \frac{{\dot {x}_{c}}^2}{4 D_1 \lambda} } \biggr)^{-1/2}, \\
\label{e8} {\dot {\bf x}_{c}}& = {\bf \xi} - \kappa
\frac{\epsilon}{8 \pi D_1^{2}} Ei \Biggl( \lambda \tau + \frac{\tau
{\dot {x}_{c}}^{2}}{4 D_1} \Biggr) {\dot {\bf x}_{c}},
\\
\label{e9} {\dot {\bf x}_{c}}& = {\bf \xi} - \kappa
\frac{\epsilon}{8 \pi^ {3/2} D_1^{5/2} \sqrt {\tau}} \Biggl(1 -
\sqrt{\pi \lambda \tau+ \pi \frac {\tau {\dot {x}_{c}}^{2}}{4 D_1} }
\Biggr) {\dot {\bf x}_{c}}.
\end{align}
Note that the the function $Ei(x)$ in Eq.(\ref{e8}) refers to the
exponential integral. In many biological cases it is found that
$\zeta = D_0 / D_1 \ll 1$ (for example $\zeta = 1/40 - 1/400$ for
Dictyostelium \cite{Hofer} and $\zeta \simeq 1/30$ for microglia
cells and for neutrophils \cite{Keshet}) and so the above triad of
equations simplify by noticing that to a first approximation we have
$\langle {\dot {x}_{c}}^2 \rangle \ll 4 D_1 \lambda$. Note that this
entails replacing the magnitude of the velocity squared ${\dot
{x}_{c}}^{2}$ in Eqs.(5--7) by its average over noise $\langle {\dot
{x}_{c}}^2 \rangle$. Then the equations are all reduced to the
Langevin form for a pure random walk, with a dimensionally-dependent
renormalized cell diffusion coefficient $D_r$ of the form:
\begin{equation}
\label{e11a} D_r = {D_0} (1 + \kappa \tilde{\epsilon}_d ) ^ {-2},
\end{equation}
where
\begin{align}
\label{e12} \tilde{\epsilon}_1 &= \frac{\epsilon(1-erf \sqrt{\lambda \tau})}{4 D_1^{3/2} \sqrt{\lambda}}, \\
\label{e13} \tilde{\epsilon}_2 &= \frac{\epsilon Ei(\lambda \tau)}{8 \pi D_1^{2} }, \\
\label{e14} \tilde{\epsilon}_3 &= \frac{\epsilon}{8 \pi^{3/2}
D_1^{5/2} \sqrt{\tau}}.
\end{align}
The expressions for $D_r$ are consistent provided they do not
invalidate the initial assumption $\langle {\dot {x}_{c}}^2 \rangle
\ll 4 D_1 \lambda$. It is easy to show that the above treatment is
justified given that the inequality $\label{e15} D_r / 2 D_1 \lambda
\delta t \ll 1$ is met, where $\delta t$ is a typical correlation
time for the cell's direction of movement. The inequality verifies
our initial approximation used in deriving Eq.(\ref{e11a}), namely
that the condition $\zeta \ll 1$ allows us to neglect the factor
${\dot {x}_{c}}^2 / 4 D_1 \lambda$ in Eqs.(5--7). The validity of
our results is also confirmed by numerical simulations. Fig.1 shows
a plot of $D_r/D_0$ versus the coupling strength $\epsilon$ for
three different ratios of $\zeta$ in one dimension ($\kappa = 1$).
\begin{figure} [h]
\includegraphics [width=3.5in] {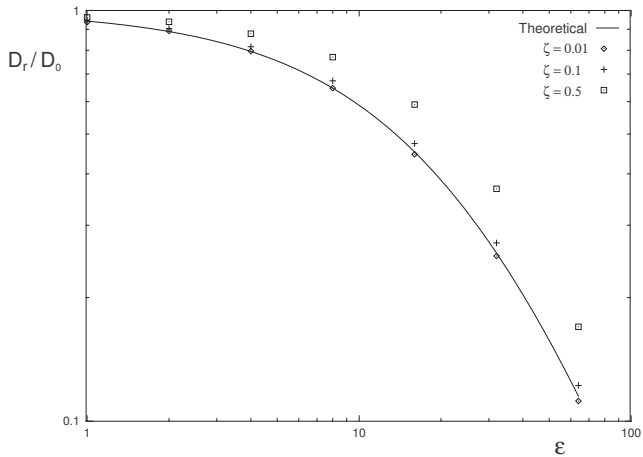}
\caption{Renormalization of the single cell diffusion coefficient in
one dimension. The parameters used are $D_1=10$, $\lambda=0.05$, and
$\delta t = 0.3$. The number of samples taken is 5 $\times$
$10^{4}$. $D_0$ is 0.1 for the circle data points, 1.0 for the plus
data points, and 5.0 for the square data points. The solid line is
the prediction from the non-perturbative method in the limit of
small $\zeta$.}
\end{figure}
Expanding the equations for $D_r$ in a power series for $\epsilon$
up to and including terms in $\epsilon^2$, we find that these
expressions agree exactly with those from first and second-order
perturbation theory in the limit of small $\zeta$ \cite{Newman}. The
advantage of the non-perturbative method over its perturbative
cousin, is its simplicity and its theoretical validity for all
coupling strengths. The non-perturbative results suggestively
indicate that for positive chemotaxis ($\kappa = 1$), for large
coupling $\epsilon$ independent of the values of $D_{0}$, $D_{1}$
and $\lambda$ (provided $\lambda > 0$) the cell's asymptotic motion
can be described by a random walk with a renormalized diffusion
coefficient. In particular we have the prediction $D_r \propto
\epsilon^{-2}$. Since $D_r$ is always positive and greater than zero
this clearly shows that self-localization due to auto-chemotaxis is
impossible in all dimensions. Applying the same methodology to
solving the case of $N$ interacting cells, one finds that contrary
to the single cell case it is not possible to decouple the equations
in such a way so as to determine an approximate equation of motion
for each cell. However it is possible to determine an equation of
motion for the center of mass of the interacting cells. In
particular one finds that if aggregation occurs then the center of
mass of the aggregate has a renormalized diffusion coefficient
\begin{equation}
\label{e69} D_R = D_{0} N^{-1} \biggl[1+ N \gamma \int\limits_{\tau
/ t}^{1} du \frac {e ^{ {-\lambda t u}}} {u^{d/2}} \biggr]^{-2}.
\end{equation}
In the limit of large coupling strength, independent of dimension
$d$, the above equation is reduced to the simple form $D_R \propto
\epsilon^{-2} N^{-3}$. The latter implies that fluctuations in the
position of the center of mass decrease as $N^{-3/2}$ (Note that in
the absence of chemotaxis i.e. $\epsilon = 0$, the fluctuations
decrease as $N^{-1/2}$, as expected). In the mean-field equations,
the center of mass corresponds to the quantity $\int d^{d} x \
{\bf{x}} \ \rho({\bf{x}},t) / \int d^{d} x \ \rho({\bf{x}},t)$. For
the case of aggregation, the latter quantity agrees with the mean
position of the center of mass obtained from the stochastic model.
However note that whereas the mean-field equations can only give
information about the average position of the center of mass of the
aggregate, the stochastic equations characterize the fluctuations
about this mean. These fluctuations may play an important role in
the fusion of two separate but close aggregates in which the number
of cells is not very large. Such a phenomenon would lead to
different temporal evolution histories (though not necessarily a
different final outcome) between the stochastic and mean-field
equations.

We now turn our attention to the case of a cell self-interacting via
negative chemotaxis i.e $\kappa = -1$. Renormalized diffusion,
Eq.(\ref{e11a}), is the cell's asymptotic behavior; this is exactly
as for positive chemotaxis, though now $D_r > D_0$. However note
that $D_r$ has a singularity when the coupling strength equals a
certain critical value given by $\tilde{\epsilon}_d = 1$. This
indicates a possible transition from renormalized diffusive motion
(for weak coupling) to a different type of motile behavior. Since we
are postulating a transition to behavior other than diffusion, the
relevant parameter to investigate is $\Lambda$ which is defined
through the mean square displacement of the cell as: $\langle
x_c^{2} \rangle \propto t^{\Lambda}$. Numerical simulations in one
dimension show that asymptotically $\Lambda = 1$ for $\epsilon < 4
D_1^{3/2} \sqrt{\lambda}$ whereas for $\epsilon > 4 D_1^{3/2}
\sqrt{\lambda}$ invariably we have $\Lambda = 2$ (Fig. 2). We shall
refer to this phase as ballistic. For $\epsilon$ very close to the
critical point we find that the system takes a very long time to
stabilize into its asymptotic limit, a feature typical of phase
transitions in physical systems \cite{Kadanoff}.
\begin{figure} [h]
\centering
\includegraphics [width=3.3in] {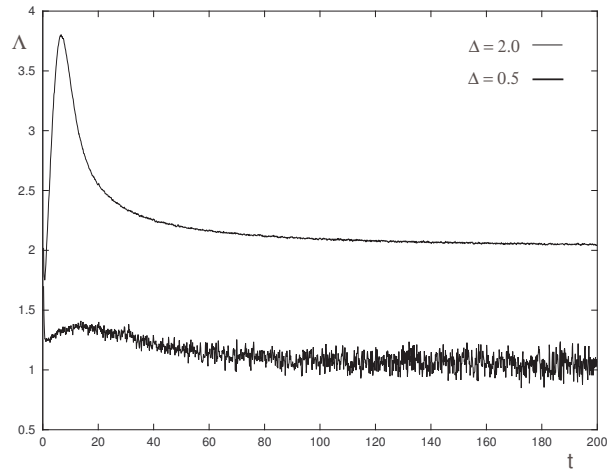}
\caption{Graph showing the asymptotic value of $\Lambda$ for two
values of the parameter $\Delta = \epsilon / 4 D_1^{3/2}
\sqrt{\lambda}$ in one dimension. For $\Delta < 1$, the asymptotic
value of $\Lambda$ is unity, while for $\Delta > 1$, $\Lambda$ takes
the value of 2. This result supports the transition predicted by
theory. $\Lambda$ is computed using the relation $\Lambda = d(\log
\langle x_c^{2} \rangle)/d(\log t)$. For the case $\Delta=2$, data
is averaged over $10^4$ samples whereas 2 $\times$ $10^5$ samples
were used for $\Delta=0.5$. The parameter values used are
$D_0=0.01$, $D_1=1$, $\lambda=0.1$ and $\delta t=0.1$. Note that
$\tau$ is chosen small enough so that it satisfies the condition
$erf \sqrt{\lambda \tau} << 1$.}
\end{figure}
It is possible to gain some understanding on the nature of the
transition by temporarily ignoring the noise vector $\xi$ in
equations Eqs.(5--7), and analyzing the then deterministic
equations. Note that ignoring the noise is plausible for the case
$\zeta \ll 1$ since this qualitatively implies that the noise term
is small compared to the velocity term in the Langevin equation
Eq.{\ref{e1}}. For positive chemotaxis ($\kappa = 1$), the only
solution in all dimensions is the trivial solution ${\dot {\bf
x}_{c}}=0$. Thus if the cell is momentarily perturbed from its
original position, it will move for a short time and then come to a
complete halt, signifying the stability of the equilibrium state.
This stability is independent of the strength of the perturbation or
the time at which the perturbation is applied as long as the
perturbation is not continuous. This result is also compatible with
the form of the renormalized diffusion coefficients derived for
positive chemotaxis i.e. in the limit of very strong coupling
(chemotaxis dominating over the noise) the cell motility becomes
very small. For negative chemotaxis ($\kappa = -1$), there exist two
real solutions: the trivial solution ${\dot {\bf x}_{c}}=0$ and a
non-zero solution obtained through algebraically solving for the
cell velocity. For $\tilde{\epsilon}_d < 1$, the only solution is
the trivial solution however for $\tilde{\epsilon}_d > 1$, both
solutions are possible. This means that for weak coupling, a cell
which is perturbed from its original position, wanders around and
eventually stops moving. However for coupling strengths larger than
a critical coupling strength if the cell is perturbed from its
original state then it will move with constant speed in the same
direction in which it was originally perturbed. In this case the
equilibrium state is unstable. Thus the zero noise analysis predicts
the observed sharp transition in $\Lambda$ at the critical coupling
$\tilde{\epsilon}_d = 1$, for small $\zeta$. The expressions for the
deterministic cell velocity ($\tilde{\epsilon}_d > 1$) obtained from
such a treatment are also found to be in good agreement with the the
root mean square cell position divided by the time, obtained from
simulations. It is interesting to note that in-vitro experiments
investigating the negative chemotaxis phase of an initially compact
aggregate of \emph{Dictyostelium}, show that the cells' displacement
is proportional to time and not to the square root of time as normal
non-chemotactic cells do \cite{Keating}. This is concordant with our
theory, since for an initially dense aggregate of cells, dispersion
forces the self-interaction of cells to take over the asymptotic
dynamics i.e. ballistic behavior is the predicted outcome. It is
notable that such behavior is not obtained from the Keller-Segel
equations (the equations referred to in this case are the
Keller-Segel equations \cite{Keller} with a negative $\alpha$
instead of a positive one, as is usually the case for positive
chemotaxis).

Concluding we have shown that (i) a single cell self-interacting via
positive chemotaxis ($D_0 \ll D_1$) performs a random walk
characterized by a renormalized diffusion coefficient $D_r > 0$.
This implies that the self-localization of a single chemotactic cell
is impossible, independent of the strength of the coupling between
the cell and the chemical field. (ii) a system of cells aggregating
via positive chemotaxis leads to an aggregate whose center of mass
performs a random walk with a renormalized diffusion coefficient.
The latter characterizes the fluctuations about the center of mass,
information not given by the mean-field model. For large coupling,
fluctuations in the aggregate center of mass decrease as $N^{-3/2}$
and thus in this regime, the differences in the temporal evolution
predicted by the stochastic and mean-field equations may not be very
large. This may explain why the mean-field models have been
successful at qualitatively modeling a number of chemotactic
phenomena. For biological cases where $\gamma$ is not large, the
fluctuations are considerably larger and thus the differences
between the two types of models may be more pronounced. (iii)
Negative chemotaxis results in either diffusive or ballistic
behavior. Whereas for chemotactic aggregation, one could argue that
the mean-field model equations (i.e. the Keller-Segel equations)
become a better description at later times, when the cell number
density becomes large, this is not the case for dispersion via
negative chemotaxis. This is borne out by our simulations. Indeed
this may apply to any system which involves cellular interactions
via negative chemotaxis (e.g. the directional control of axonal
growth in the wiring of the nervous system during embryogenesis
\cite{Painter}). 

It is a pleasure to thank Timothy Newman for interesting
discussions. The author gratefully acknowledges partial support from the NSF (DEB-0328267, IOB-0540680).

\end{document}